\documentclass[preprint2]{aastex}

\begin{document}

\def\la{Ly$\alpha$}
\def\hb{H$\beta$}
\def\arcmin              {$^{\prime}$}
\def\arcm                {$^{\prime}$}
\def\arcsec              {$^{\prime\prime}$}
\def\arcs                {$^{\prime\prime}$}
\def\kms                 {km\thinspace s$^{-1}$}
\def\cms                 {cm$^{-2}$}
\def\etal{{\it et al. }}
\def\aa{{\rm A$\,$\&$\,$A}}            
\def\araa{{\rm ARA$\,$\&$\,$A}}                
\def\aa
\documentstyle[psfig,referee]{mn}
\def\la{Ly$\alpha$}
\def\hb{H$\beta$}
\def\arcmin              {$^{\prime}$}
\def\arcm                {$^{\prime}$}
\def\arcsec              {$^{\prime\prime}$}
\def\arcs                {$^{\prime\prime}$}
\def\kms                 {km\thinspace s$^{-1}$}
\def\cms                 {cm$^{-2}$}
\def\etal{{\it et al. }}
\def\aa{{\rm A$\,$\&$\,$A}}            
\def\araa{{\rm ARA$\,$\&$\,$A}}                
\def\aar{{\rm A$\,$\&$\,$AR}}          
\def\aas{{\rm A$\,$\&$\,$AS}}          
\def\apj{{\rm ApJ}}                    
\def\apjs{{\rm ApJS}}                  
\def\apjl{{\rm ApJ Let}}               
\def\aj{{\rm AJ}}                      
\def\sva{{\rm SvA}}                    
\def\pasp{{\rm PASP}}                  
\def\pasj{{\rm PASJ}}                  
\def\mnras{{\rm MNRAS}}                        
\def\kmsmpc              {km\thinspace s$^{-1}$\thinspace Mpc$^{-1}$}
\def\Msol{\thinspace\hbox{$\hbox{M}_{\odot}$}}
\def\Zsol{\thinspace\hbox{$\hbox{Z}_{\odot}$}}
\def\sol{\thinspace\hbox{$_{\odot}$ }}
\def\deg{\hbox{$^\circ$}}
\def\kpc{\thinspace\hbox{kpc}}
\def\ojo{\fbox{\bf !`$\odot$j$\odot$!}}

\newcommand{\der}[2]  { \frac{{\rm d}#1}{{\rm d}#2} }
\newcommand{\derp}[2] { \frac{\partial #1}{\partial #2} }
\newcommand{\dif}     {{\rm d}}
\newcommand{\difp}    {\partial}
\renewcommand{\thefootnote}{\fnsymbol{footnote}}

\title{On the recent history of star formation in the BCD galaxy VII Zw403.}

\author{S. Silich \altaffilmark{1,2}, 
G. Tenorio-Tagle \altaffilmark{1,}\footnote{Visiting Professor at IoA, 
Cambridge},
        C. Mu\~noz-Tu\~n\'on \altaffilmark{3} and 
        L. M. Cairos \altaffilmark{3,4}}
\altaffiltext{1}{Instituto Nacional de Astrof\'\i sica Optica y Electr\'onica, 
                 AP 51, 72000 Puebla, M\'exico}
\altaffiltext{2}{Main Astronomical Observatory National Academy of Sciences of
                 Ukraine, 03680, Kiev-127, Golosiiv, Ukraine}
\altaffiltext{3}{Instituto de Astrof\'\i sica de Canarias, E 38200 La Laguna, 
                 Tenerife, Spain}
\altaffiltext{4} {Departamento de Astronom\'\i a, Universidad de Chile,
                 Casilla 36-D, Santiago, Chile}


\begin{abstract}
{Here we attempt to infer the recent history of star formation in the BCD 
galaxy VII Zw403, based on an analysis that accounts for the dynamics of 
the remnant generated either by an instantaneous burst or by a continuous star
formation event. The models are restricted by the size of the diffuse 
X-ray emitting region, the H$_{\alpha}$ luminosity from the
star-forming region and the superbubble diffuse X-ray luminosity. 

We have re-observed VII Zw403 with a better sensitivity corresponding
to the threshold H$_{\alpha}$ flux $8.15 \times 10^{-17}$ erg
cm$^{-2}$ s$^{-1}$. The total H$_{\alpha}$ luminosity derived from
our data is much larger than reported before, and presents a variety
of ionized filaments and incomplete shells superimposed on the
diffuse H$_{\alpha}$ emission. This result has a profound impact on 
the predicted properties of the starburst blown superbubble.
Numerical calculations based on the HST H$_{\alpha}$ data, 
predict two different scenarios of star formation able to match
simultaneously all observed parameters. These are an instantaneous burst 
of star formation with a total mass of $5 \times 10^5$ M$_{\odot}$ and 
a star-forming event with a constant SFR = $4 \times 10^{-3}$ 
M$_{\odot}$ yr$^{-1}$, which lasts for 35 Myrs. The numerical
calculations based on the energy input rate derived from our
observations predict a short episode of star formation lasting less
than 10 Myrs with a total star cluster mass $\sim (1 - 3) 
\times 10^6$ M$_{\odot}$.
However, the five main star-forming knots are sufficiently
distant to form a coherent shell in a short time scale, and still
keep their energies blocked within local, spatially separated bubbles.
The X-ray luminosities of these is here shown to be consistent with
the ROSAT PSPC diffuse X-ray emission.}

\end{abstract}


\section{Introduction}

It has recently been recognized that the star formation activity in galaxies 
is very irregular in time, and many examples of major burst episodes 
exhibit an extremely high star formation rate concentrated in well localized 
space regions (Terlevich 1996). It is also now well known that starbursts 
(SBs) cause an emission that dominates the entire host galaxy luminosity and 
their mechanical energy input rate is expected to cause major structural 
changes in the surrounding  interstellar medium (ISM). In this respect 
it has become of great interest to study the properties of the resultant  
large-scale expanding superbubbles which, powered by the violently injected 
newly processed matter, establish the time scale for mixing with the ISM
(Tenorio-Tagle 1996, Silich et al. 2001). In extreme cases, the superbubbles 
are thought to break out of the galactic discs leading to an effective mass 
and energy transport into the low density halos or even into the 
intergalactic medium via a superwind (Heckman \etal 1990). 

Starbursts in the local universe are also assumed to be good representatives 
of the star-forming activity at high redshifts. This concept defines their 
cosmological interest as key laboratories for studying the ISM, the transport 
of supernovae processed metals, as well as the chemical evolution of galaxies 
and of the intergalactic medium. The resulting structure in the ISM
due to mechanical energy injected by SBs is very similar to the interstellar 
wind-blown bubbles around single massive stars (see Weaver \etal 1977 for 
their four zone model), although the much larger energy input rate in SBs 
leads rapidly to much larger scales. Hydrodynamical simulations (see 
Tenorio-Tagle \& Bodenheimer 1988, Bisnovatyi-Kogan \& Silich 1995 and 
references therein; Suchkov \etal 1994; Silich \& Tenorio-Tagle 1998; 
D'Ercole \& Brighenti 1999; Strickland \& Stevens 2000) currently include  
differential galactic rotation, radiative cooling, strong density gradients 
between the disk and the halo and thus are able to follow the moment of 
breakout, as well as the fragmentation of the expanding outer shell via 
Rayleigh - Taylor instabilities and the venting of the superbubble hot 
interior gas, either into the intergalactic space, or into the host galaxy 
halo (the blowout phenomenon).
  
Most of the up to-date simulations have been performed under the assumption 
of a constant energy deposition rate, as expected from an
instantaneous burst model.   
However, studies of the stellar population in OB associations related 
to young ($\tau_{OB} < 10$ Myr) Large Magellanic Cloud (LMC) bubbles 
(see  Oey \& Smedley, 1998 and references therein) have demonstrated that 
``realistic'' energy input rates are very different from the assumed constant 
energy input rates used in numerical simulations. Thus the instantaneous burst 
assumptions may not be applicable to all cases.
Here we attempt to establish a method of comparison between the theory of 
superbubbles and the observations of remnants produced by massive star
formation in galaxies. Two possible modes of star formation, instantaneous and 
extended bursts, are taken into consideration. For both cases the mechanical 
luminosity, ultraviolet photon output, mass returned to the ISM, and the 
fraction of each in metals, all as a function of time,  are estimated. 
On the other hand we have the observed parameters: the H$_{\alpha}$ or 
H$_{\beta}$ luminosity which can be directly related to the SFR under the 
assumption that all photons are used up in the ionized region. One can also 
estimate the size and luminosity of the X-ray remnant. In some cases the 
remnants may have slowed down sufficiently to display their outer 
expanding shells, either in the optical or in HI observations, or both, giving 
further information about the size, expansion speed, and mass behind the 
outer shock. A comparison with the theory also requires some 
preconception of the galaxy's ISM, which can as a first approximation 
be derived from HI observations and the inferred dynamical mass.
However, one also needs to make an assumption about the fraction of this gas 
locked up in dense clouds and immersed into a less dense  ISM background.
   
With the aim of establishing a method to confront theory with observations 
that may lead us to infer the recent star formation history of galaxies, 
the blue compact dwarf (BCD) VII Zw403 is thoroughly analyzed. Section
2 derives the main properties of coeval and extended star formation
modes. Section 3 summarizes the main observational properties of our 
target. Section 4 presents a summary of the hydrodynamical
calculations aimed at matching the observed properties. The results
of the calculations and our main findings are discussed in Section 5.

\section{Star formation}

Following the observational evidence in support of the concept of a continuous
star formation spread over 10 - 20 Myr or more (Herbst \& Miller 1982, 
Stahler 1985, Coziol et al. 2001), Shull \& Saken (1995) have discussed five 
possible tracks of the star formation rate (SFR) time evolution, including 
both the time span of star formation, and possible time variations in the 
initial mass function (IMF). Here we examine one of their cases: a 
constant SFR spread over a finite time interval $\tau_{SB}$. We
compare its intrinsic properties with the standard instantaneous 
model. The simple model discussed below can be generalized to more 
sophisticated cases with a time dependent SFR and a variable IMF. 
    
\subsection{The properties of starbursts}

Let us  assume an instantaneous burst characterized by an IMF with an exponent 
$\alpha$,
\begin{equation}
      \label{eq.1}
f(m) = \frac{(\alpha-2) M_{SB}}{M_{low}^{2-\alpha} - M_{up}^{2-\alpha}} 
m^{-\alpha},
\end{equation}
where $M_{SB}$ is the total mass of the SB cluster, and $M_{low}$ and $M_{up}$
are the lower and upper cut-off masses, respectively. From this  one can 
calculate the expected energy and mass input rates. 

\subsubsection{ The mechanical energy input rate.}

The mechanical energy deposition from a starburst should include both the
energy injection via stellar winds and from supernova explosions.
Here the early mechanical energy injection (before the first supernova 
explosion at $t$ = $t_{sn}$) has been approximated by a constant value $L_w$, 
which adds the contribution of all individual massive stars in the cluster. 
This value can easily be scaled from the template of Leitherer \& Heckman, 
1995 (hereafter LH95) for a starburst with a total mass of 10$^6$ M$_{\odot}$
within the mass range $M_{low}$ = 1 M$_{\odot}$ and $M_{up}$ = 100 M$_{\odot}$
\begin{equation}
      \label{eq.2}
L_{SB} = L_{LH} \frac{M_{SB}}{M_{LH}} = Const, \qquad t \le t_{sn}, 
\end{equation}
where  $M_{SB}$ is the total stellar mass considered, 
$M_{LH}$ = 10$^6$ M$_{\odot}$ and $L_{LH}$, following LH95 was set equal 
to $10^{39}$ erg s$^{-1}$.

After a time ($t > t_{sn}$) the energy released via supernova explosions
dominates the SB mechanical energy deposition, and then the energy released 
by the SB as a function of time is defined by the expression
\begin{eqnarray}
      \label{eq.3}
      & & \hspace{-0.9cm} \nonumber
E_{SB}(t) = \int_{M(t)}^{M_{up}} E_{SN} f(m) {\rm d}m =
      \\[0.2cm]
      & & \hspace{-0.9cm}
           \frac{(\alpha - 2)}{(\alpha - 1)} 
           \frac{M_{SB} E_{SN}}{M_{low}^{2-\alpha} - M_{up}^{2-\alpha}}
           \left(M^{1-\alpha}(t) - M_{up}^{1-\alpha}\right), 
\end{eqnarray} 
where $M(t)$ is the mass of the stars exploding as supernovae after an 
evolutionary time $t$. For simplicity, the energy released per supernova 
($E_{SN}$) is assumed to be independent of the progenitor mass and equal 
to 10$^{51}$ erg.

The energy input rate from the SB at the supernova dominated stage is then 
\begin{eqnarray}
      \label{eq.4}
      & & \hspace{-0.9cm} \nonumber
L_{SB} = \der{E_{SB}(t)}{t} = 
      \\[0.2cm]
      & & \hspace{-0.9cm}
    - \frac{(\alpha - 2) M_{SB} E_{SN} M^{- \alpha}(t)}
      {M_{low}^{2-\alpha} - M_{up}^{2-\alpha}} \der{M}{t},
      \quad t > t_{sn}. 
\end{eqnarray}
The lifetime of the massive stars can be inferred from  the approximations 
of Chiosi et al. (1978) and Stothers (1972). Then the  function $M(t)$ has  
the form:
\begin{equation}
      \label{eq.5}
\frac{M(t)}{10M_{\odot}}  = \left\{
\begin{array}{lcl}
(9 \times 10^6 / t)^2 ; \hfill \ 
            30M_{\odot} \le M \le 100M_{\odot}
\\ [0.2cm]
(3 \times 10^7 / t)^{5/8} ; \hfill \  
          7M_{\odot} \le M \le 30M_{\odot}.
\end{array}
\right.
\end{equation}
Substituting the time derivative of $M(t)$ into equation (\ref{eq.4}) one 
obtains
\begin{equation}
      \label{eq.6}
L_{SB}  = \left\{
\begin{array}{lcl}
  \frac{2 (\alpha - 2) M_{SB} E_{SN}}
  {M_{low}^{2-\alpha} - M_{up}^{2-\alpha}} \frac{M^{1 - \alpha}(t)}{t} ,   
          \hspace{0.1cm}       t_{sn} \le t \le t_{30}
\\ [0.2cm]
  \frac{5}{8} \frac{(\alpha - 2) M_{SB} E_{SN}}
  {M_{low}^{2-\alpha} - M_{up}^{2-\alpha}} \frac{M^{1 - \alpha}(t)}{t} , 
          \hspace{0.1cm}       t_{30} \le t \le t_{7}
\end{array}
\right.
\end{equation}
where $t_{30}$ and $t_{7}$ are the stellar lifetimes for a 30 M\sol and 
a 7 M\sol star, respectively.

\subsubsection{Mass deposition}

Before the first supernova explosion, the mass injected by a star cluster
results from individual stellar winds calculated as
\begin{equation}
      \label{eq.7}
M_{ej}(t) = \int_{0}^{t} 2 L_{SB} / V_w^2 dt, \quad t \le t_{sn},
\end{equation}
where the collective wind terminal velocity is assumed to be 
$V_w = 2000$ km s$^{-1}$. Afterwards,  the  mass injection is dominated by 
supernovae and one can  neglect the contribution from stellar winds. During  
the supernova dominated stage, the total ejected mass as a function of time 
may be approximated by (see Silich et al. 2001):
\begin{equation}
      \label{eq.8}
M_{ej}(t) =
M_{SB} \frac{M(t)^{2-\alpha} - M_{up}^{2-\alpha}}
                     {M_{low}^{2-\alpha} - M_{up}^{2-\alpha}} ,
                     \quad t > t_{sn}.
\end{equation}

Following  Silich et al. (2001) we assume that the gas ejected by 
supernovae includes all the newly synthesized metals, and thus the
mass in ejected metals as a function of time is
\begin{eqnarray}
      \label{eq.9}
      & & \hspace{-0.9cm} \nonumber
M_{metals} = 
      \\[0.2cm]
      & & \hspace{-0.9cm}
\frac{(\alpha -2) M_{SB}}{M_{low}^{2-\alpha} - M_{up}^{2-\alpha}}
\int_{M(t)}^{M_{up}} Y_{metals}(m) m^{-\alpha} {\rm d}m ,
\end{eqnarray}
where $Y_{metals}$ is a particular element yield. One can then use, 
for example,  
oxygen as tracer of the  metalicity caused in the interior of superbubbles. 
In such a case the oxygen yield can be approximated by analytical fits to 
the stellar evolutionary tracks of Maeder (1992) and Woosley \etal (1993), 
which account for the mass loss due to stellar winds (see Silich et al. 2001).
The mean metalicity inside a superbubble  is then defined by the ratio
of the wind and supernovae ejected oxygen to the total mass within the
superbubble $M_{in} = M_{ej} + M_{ev}$, which includes also the mass thermally 
evaporated from a cold outer shell  ($M_{ev}$):
\begin{equation}
      \label{eq.10}
Z_{O} = \frac{M_{ej}(O)/Z_{\odot}(O) + Z_{ISM} M_{ev}}
                        {M_{in}} ,
\end{equation}
where $Z_{ISM}$ and $Z_{O}$ are both in the solar units, 
$Z_{\odot}(O) = 0.0083$ (Grevesse et al. 1996). 

\subsubsection{The $UV$ output}

For a coeval star cluster, the $UV$ photon  production rate remains 
almost constant until the most massive star in the cluster begins to
move away from the main sequence, soon becoming a supernova. From this 
moment onward, the $UV$ flux rapidly decays (Beltrametti \etal 1982).
We approximate this evolutionary track by the power function  
\begin{equation}
      \label{eq.11}
N_{UV}  = \left\{
\begin{array}{lcl}
          N^0 \frac{M_{SB}}{M_{LH}},
          \qquad \qquad  \quad    0  \le t \le t_{sn}
\\ [0.2cm]
          N^0 \frac{M_{SB}}{M_{LH}}(t_{sn} / t)^5 ,
          \qquad         t > t_{sn} .
\end{array}
\right.
\end{equation}
The initial flux ($N^0$) was normalized to the 10$^6$ M$_\odot$ standard 
model of LH95, and was taken to be 9 $\times 10^{52}$ 
photons s$^{-1}$. If all $UV$ photons are trapped within a star forming
region, then the H$_{\alpha}$ expected luminosity $L_{H_{\alpha}}$ 
follows from a simple transformation (LH95):
$L_{H_{\alpha}} = 1.36 \times 10^{-12} N_{UV}$ erg s$^{-1}$.

\subsection{Continuous star formation}

If the star formation process is not coeval but is instead spread over a 
time $\tau_{SB}$, one can approximate it by a series of $N_{tot}$ 
instantaneous sequential mini bursts separated by a $\Delta \tau = 
\tau_{SB} / N_{tot}$. One can further assume that all mini bursts would have 
the same mass $M_i = M_{SB} / N_{tot}$ and IMFs and would evolve 
independently according to their own clocks set upon formation:
\begin{equation}
      \label{eq.13}
t_i = t - (i - 1) \Delta \tau ,
\end{equation}
where $t$ is an evolutionary time, and $i$ is the mini burst
number. Then the number of mini bursts at any given time $t$ is
\begin{equation}
      \label{eq.14}
n(t)  = \left\{
\begin{array}{lcl}
          Int(t / \Delta \tau) + 1 ,
          \quad   if   \quad  t \le \tau_{SB}
\\ [0.2cm]
          N_{tot} ,
          \quad   if   \quad  t > \tau_{SB} .
\end{array}
\right.
\end{equation}
The cumulative intrinsic properties (mechanical energy input rate, mass 
injection and the number of $UV$ photons or the H$_{\alpha}$ luminosity) 
can be found by simply adding the mini burst parameters: 
\begin{eqnarray}
      \label{eq.15}
      & & \hspace{-0.5cm}
L_{SB} = \sum_{i=1}^{i=n(t)} L_i ,
      \\[0.2cm]
      & & \hspace{-0.5cm}
M_{ej} = \sum_{i=1}^{i=n(t)} M_{ej,i} ,
      \\[0.2cm]
      & & \hspace{-0.5cm}
N_{UV} = \sum_{i=1}^{i=n(t)} N_{UV,i} .
\end{eqnarray}

Note that the input from each mini burst  should be set to zero whenever
their individual evolutionary time $t_i$ exceeds the lifetime of the less
massive star which can explode as supernova; that is once  $t_i$ becomes 
larger than  their intrinsic $t_7$.

The global energy properties of star formation events derived from this simple
model are in a good agreement with the two extreme cases (instantaneous and
continuous star formation) of LH95 and are presented in Figure 1.
For comparison a value of $\alpha$ = 2.35 has been assumed as well as a 
stellar range of masses between 1 M$_\odot$ and 100 M$_\odot$. 

Figure 1a compares the mechanical energy input rates derived for a
10$^6$ M$_{\odot}$ stellar cluster formed during a $\tau_{SB}$ = 1 Myr 
(hereafter the instantaneous 
burst), 20 Myr and 50 Myr. The solid line shows the rapid rise in the coeval 
case, which reaches a maximum immediately after the onset of supernova 
explosions, and then drops to values $\sim$ 10$^{40.2}$ erg s$^{-1}$ to 
remain almost constant up to the end of the supernova activity 
($\sim$ 50 Myr). For the $\tau_{SB}$ = 20 Myr event, the maximum energy 
deposition is delayed in time, and converges with the 1 Myr values after 
approximately 25 Myr. The energy deposition rate for $\tau$ = 50 Myr star
forming episode suffers an even longer delay, to reach values similar to the 
maximum input from the previous two models, and  then slowly decays without 
reaching a uniform value throughout the 100 Myr of its evolution.

The production rate of $UV$ photons (see Figure 1b) is highly dependent on 
the star formation time considered. For the 1 Myr 
starburst the $UV$ flux reaches  its maximum value 
($\sim$ 9 $\times 10^{52}$) at t = $\tau_{SB}$, and remains constant during 
the first 4 Myr. It then  falls very steeply, reaching before 10 Myr of 
evolution, values more than two orders of magnitude smaller than its maximum 
value. The 20 Myr and 50 Myr clusters acquire their uniform constant value 
after 4 Myr of evolution, and then upon completion of their star formation 
phase, their $UV$ photon output also rapidly decays with time.

The rate at which mass is ejected by the massive star cluster is shown in 
Figure 1c. The total ejected mass ($\sim 40 \%$ of the star cluster mass) 
and the total oxygen mass released by supernovae (about 2$\%$ of the total 
mass turned into stars) is clearly the same at the end of the evolution in 
the three cases considered. There are however substantial delays for 
the larger the value of $\tau_{SB}$.

\section{The main observational properties of VII Zw403}

VII Zw403 is a BCD galaxy considered in recent years in many 
discussions related to star formation histories and possible
impact of dwarf systems on the surrounding intergalactic medium.
Although Tully et al. (1981) proposed that the galaxy might be a member of 
M81 group, more recent distance determinations (Lynds et al. 1998) locate 
it about 1.5 Mpc further away at 4.5 Mpc distance. Therefore it can be 
considered an isolated galaxy (Schulte-Ladbeck et al., 1999).

The X-ray observations (Papaderos et al. 1994; Fourniol 1997) added more 
interest to the system, as they revealed an extended kpc-scale region of 
diffuse X-ray emission. These observations account for approximately 
85\% of the total X-ray luminosity (L$_{X,total} \approx 2.3 \times 
10^{38}$ erg s$^{-1}$) from the central unresolved core, while the 
remaining 15\% is spread 
over the central kpc-scale region. The high-resolution data reported
by Lira et al. (2000) revealed a strong point X-ray source displaced
to the west with respect to the main H$_{\alpha}$ emission. This is
most probably a powerful (L$_X \sim 10^{38}$ erg s$^{-1}$) binary
system. However, the kpc-scale diffuse component was not confirmed
by these observations, possibly due to the low sensitivity of the
HRI.

The total dynamical mass of the galaxy is $2 \times 10^8$ \Msol, with
approximately 20\% or $4 \times 10^7$ \Msol \, constituting the neutral 
hydrogen
phase (Thuan \& Martin, 1981). VII Zw403 does not exhibit a regular rotation, 
but rather turbulent random motions of individual neutral hydrogen clumps 
with $\Delta V_t \approx 30$ km s$^{-1}$ (Thuan \& Martin, 1981). This 
value is taken as the velocity dispersion which supports the ISM 
against the self gravity of the galaxy. The extension of the neutral hydrogen
halo R$_{ISM}$ is not well known, but if the average ratio of the HI size to 
the optical size is accepted to be f = 2.4 (Thuan \& Martin, 1981), the 
HI halo radius  R$_{HI}$ is about  1.9 kpc. A similar result is obtained  
if one  assumes that the galaxy boundary occurs at the location where the 
escape velocity drops below the ISM random velocity $\Delta V_t$.

Lynds et al. (1998) showed  that the stellar population of VII Zw403 
contains not only young blue main-sequence stars, but also a much
older evolved population that can be traced up to 1-2 Gyrs back in time. 
They also studied the five centrally concentrated associations of young stars,
with ages smaller than 10 Myr, and that in  H$_{\alpha}$ emission appear 
to be surrounded by local small shells with radii $\sim 100$ pc. 
Schulte-Ladbeck et al. (1999) provided a near infrared single star 
photometry of the galaxy and confirmed that compact, young
star-forming regions are embedded into a much older, low surface 
brightness halo.

\subsection{New observations}

We present here new narrow-band images of VII Zw403 centered on the 
$H_{\alpha}$ line and on the adjacent continuum (Figure 2) which were taken 
in August 1997 at the 2.2-m 
telescope of the German--Spanish Astronomical Observatory at Calar Alto 
(Almer\'{\i}a, Spain). The instrumentation consisted of the {\em Calar Alto 
Faint Object Spectrograph} (CAFOS) and a $2048 \times 2048$ SiTe CCD chip, 
with a pixel size of 0.53 arcsec, and an un-vignetted circular field of view 
of about 11 arcmin in diameter (14.4 kpc at the accepted distance to the 
galaxy of 4.5 Mpc). The averaged seeing was 1.5 arcsec. 

The image reduction was conducted using standard procedures available in 
IRAF. Each image was corrected for bias using an average  bias frame,
and was flattened by dividing by a mean twilight flat field image. After,
they were registered (for each filter we took a set of dithered
exposures) and combined to obtain the final frame,  with cosmic rays removed 
and bad pixels cleaned. The average sky level was estimated by computing the 
mean value within various boxes surrounding the object, and subtracted out as 
a constant. Flux calibration was done through the observation of 
spectrophotometric stars from Oke (1990). For more details about data 
reduction and calibration refer to Cair\'os \etal (2001). We calculated
the integrated H$_{\alpha}$ flux out to the limiting isophote with a level
equal to $2.5 \times rms$ of the background. In our case this threshold is 
$8.15 \times 10^{-17}$ erg s$^{-1}$ cm$^{-2}$. The H$_{\alpha}$ flux
was corrected for Galactic extinction following Burstein \& Heiles (1984).
No correction for internal extinction was performed.

Our H$_{\alpha}$ image shown in Figure 2 displays a variety of shapes and 
structures which deserve discussion.
Besides the starburts knots, labeled 1 - 5 after Lynds \etal (1998), we 
detect a much fainter and extended well structured diffuse emission. 
The largest obvious feature is a broken to the SW shell-like structure,
250 pc in radius, which we associate with knot 4. Knot 2 is 
almost overlapping in projection along its Eastern rim. 

On top of features and structures which appear to build a 
network of connecting knots, there is down to our limiting flux a
diffuse-smooth emission which engulfs them all and most certainly 
results from photons leaking out of the main emitting knots.

The total H$_{\alpha}$ luminosity is higher than reported
previously. Clearly our limiting flux is much lower than that of 
Lynds \etal (their Figure 2). Our data recovers the low intensity 
H$_{\alpha}$ emission and thus provides a more precise estimate 
of the total H$_{\alpha}$ luminosity. 

The spatial resolution of the HST image (Lynds \etal, 1998)
also allows for the  detection of a much smaller H$_{\alpha}$ mini-bubble 
(cocoon), associated with knot 1. This bubble has similar energy to that 
inferred for the large superbubble reported above, however it seems
much less evolved. The energy dumped by the cluster energizing  knot 1 
has not been able to sweep and displace so efficiently the
surrounding ISM out of which it formed.

The above result has a profound impact on the modeling of starbursts.
Clearly during the first few Myr of evolution every center of star 
formation disperses its high density parental cloud. It 
is only after this task has been completed that the various star
forming centers that compose a starburst would be able to 
jointly build up a large-scale superbubble.

The main observational properties of VII Zw403 from the above mentioned 
literature and our own observations are summarized in the Table 1.

\section{A numerical model for VII Zw403}

A successful model has to be able to match simultaneously all observed 
parameters. In the case of VII Zw403 these are: the size ($R_{sh} \sim$ 1 kpc)
of the diffuse X-ray emitting region, the H$\alpha$ 
and the X-ray (L$_x \sim 3 \times 10^{37}$ erg s$^{-1}$) luminosities. 
Note that Lynds \etal (1998) give a total value L$_{H\alpha} 
\sim 1.8 \times 10^{39}$ erg s$^{-1}$ which is smaller than the
total flux detected by our observations 
(L$_{H_{\alpha}} \sim 4.7 \times 10^{40}$ erg s$^{-1}$). 
We have used both of these
values in our search of a successful model of VII Zw403. Note also that 
the values derived from X-ray observations are doubtful and require 
confirmation (Bomans, 2001). VII Zw403 presents various nuclear
centers of star formation, each of which have managed to locally structure 
the ISM, and thus present a number of loops or broken shells 
with radii $\sim 100$ pc. Another important fact is that available 
observations display neither an HI nor an HII or X-ray large-scale
shell surrounding the diffuse X-ray emission. This fact 
can be ascribed to a remnant evolving along its quasi-adiabatic phase, or 
to a neutral hydrogen shell brightness that falls below the detection limit.

All considered models assume that the observed HI mass occupies a 
smooth low density disc-halo distribution, although  an important fraction of
it ($20\% < f_c < 95\%$) is in a dense cloud component. 
This has a major impact on the evolution of remnants as it affects
both the time required to reach a given size, as well as the resultant
X-ray luminosity produced within the superbubble. Our models also
assume that the ISM is supported in hydrostatic equilibrium in the
galaxy gravitational field by the turbulent gas pressure produced by
the random gas motions with an effective velocity dispersion 
$\Delta V_t$ (see Tomisaka \& Bregman, 1993; Silich \& Tenorio-Tagle,
1998). An example of the resultant initial ISM smooth component
density distribution is shown in Figure 3.

The calculations were carried out with our 3D Lagrangian code, which accounts 
for the enrichment of the hot superbubble interior by the metals ejected via 
supernova explosions (Silich et al. 2001). Using oxygen as tracer, we derived 
the time dependent superbubble interior gas metalicity, which was then used in 
our calculations of the diffuse X-ray emission. As there are no firm 
observational restrictions on the VII Zw403 star formation timescale, we have 
examined different scenarios of the recent star formation activity in this 
galaxy. Two sets of models have been considered: an instantaneous burst, and 
an extended star formation episode. We have associated an instantaneous burst 
with the 1 Myr event and calculated the SB parameters following the
prescriptions of section 2. 
In all coeval models considered, the energy input rate has been 
derived from the assumed mass of the star forming cluster. In the
extended star formation scenarios we used instead the observed 
H$_{\alpha}$ luminosity to derived the star formation rate from
our continuous star formation model (see section 2). 
This SFR was then transformed into a total star cluster mass and
energy input rate, for the various assumed values of $\tau_{SB}$, as it
was considered in section 2. A Salpeter initial mass function, 
with upper and lower cutoff masses 1 and 100 M$_{\odot}$ respectively,
was assumed for all models.

We then calculated the time $\tau_{dyn}$ that superbubbles need 
to reach the observed 1 kpc diffuse X-ray radius. A value of $\tau_{dyn}$ 
implies an age from which one can calculate the central H$_{\alpha}$ 
luminosity and the integrated superbubble diffuse X-ray emission for every 
model. The 
results of the calculations for instantaneous burst models are summarized in 
Table 2. The various models are labeled with two separate indices that 
represent the logarithm of the SB mass (also indicated in column 2) and 
the fraction of the ISM assumed to be in the smooth component respectively. 
The preceding letter ``I'' indicates that all of these are 
instantaneous burst models. Four different star cluster masses have 
been considered: $10^5$, $5 \times 10^5$, $10^6$ and $10^7$\Msol. Column 3 
indicates the fraction of the ISM mass assumed to be stored in 
clouds. The calculated dynamical time $\tau_{dyn}$ is presented in  
column 4. After this time the shock wave has expanded to 1kpc radius, and 
the relevant parameters from the model can be compared with the observed 
values. The calculated number of $UV$ photons, the H$_{\alpha}$ 
luminosity (under assumption that all $UV$ photons are trapped within a SB 
region), and the resultant superbubble diffuse X-ray luminosity, are presented 
in columns 5, 6 and 7 respectively. Column 8 indicates whether the
superbubbles evolve along a quasi-adiabatic track (A), or have made a 
transition to a radiative phase (R) before reaching the 1 kpc radius.
The two possible modes of expansion imply different properties for the
swept-up matter: either material is still too hot to radiate effectively
and thus occupies the outer 20\% of the remnant volume, or it 
has collapsed into a thin and cold outer shell, while being exposed to
the ionizing radiation.

The second set of models assumes an extended phase of star formation.
The results of the calculations for 
these models are summarized in Table 3. The various models are labeled 
again with two separate indices that now represent the star formation time
$\tau_{SB}$ (in Myr) and the fraction of the ISM in the smooth component. 
The letter ``C'' indicates the continuous star forming mode, for which
episodes lasting 5, 10, 20 and 40 Myr have been assumed. In all these models
we assumed the same SFR ($4 \times 10^{-3}$ \Msol \, yr$^{-1}$). Column 2
indicates the total stellar mass, and columns 3 to 8
list the same variables considered in Table 2. In all of these 
models we assumed that the photons currently produced by the stellar 
clusters are all used up in reestablishing the ionization of the central 
HII region. This fact is supported by the large HI mass present in VII Zw403.
 
The resulting H$_{\alpha}$ luminosity, and the time dependent mechanical 
energy input rate, are shown in Figures 4a and b respectively.

\section{Discussion}

Table 2 shows that both low (10$^5$\Msol) and high ($10^7$\Msol) mass SB 
models are completely inconsistent with the VII Zw403 parameters
derived from Lynds et al. (1998) data. 
Indeed, the low mass models lead to a very slow expansion speed, which makes 
them reach the 1kpc size only after 10 Myr, even in the lowest density case 
with a cloud mass filling factor f$_c$ = 90\% of the observed HI mass. After 
this time the most massive stars have left the main sequence and exploded as 
supernovae,  causing a fast drop in the number of emitted $UV$ photons 
and consequently in the H$_{\alpha}$ luminosity of the associated HII region. 
This luminosity in all low SB mass cases is several orders of
magnitude below the 
currently observed value. The high mass models on the other hand, are too 
energetic, produce an exceedingly high X-ray emission, and an overwhelming
$UV$ photons flux when the superbubble radius reaches 1kpc.

The intermediate mass SBs ($5 \times 10^5 - 10^6 \Msol$) are in better 
agreement with observations. However, for the emitted number of $UV$ photons 
to be consistent with the value derived from the observed H$_{\alpha}$ 
luminosity, the smooth component of the ISM has to contain only a small 
($\sim 5\%$) fraction of the observed HI mass. The shock wave blown by 
coherent SN explosions into this low density medium remains adiabatic when it 
reaches a 1 kpc radius (see Table 2). Therefore, in these models 
(I5.7\_5, I6.0\_5) an essential fraction of the X-ray emission 
($\sim 80\%$ in the model I5.7\_5 and $\sim 50\%$ in the model I6.0\_5) 
arises from the outer adiabatic shell of swept-up matter that should occupy 
the outer $\le 20\%$ of the remnant volume. However, this is not resolved in 
the available X-ray maps.

From the results in Table 3, it is clear that for a reasonable agreement
between the observed parameters of VII Zw403 and a continuous star formation
scenario, the episode of star formation should last no less than the 
dynamical time required for the superbubble to reach the 1 kpc radius. 

The best continuous star-forming model (C40\_70), with a 
SFR = $4 \times 10^{-3}$ \Msol \, yr$^{-1}$, $\tau_{SB} = 40$ Myr, and 30\% 
of the total ISM mass in the dense cloud component, requires about 35 Myr for 
the outer shock to reach a radius $\sim$ 1 kpc. In this model an extended 
gaseous halo cannot completely prevent gas loss from the galaxy and allows a 
final shell speed slightly in excess of the escape velocity.
The H$_{\alpha}$ luminosity is not an issue in this case as it is exactly the 
amount used to derive the constant SFR. The time evolution of the superbubble
diffuse X-ray emission is shown in Figure 5 for different ISM models. 
Initially the X-ray luminosity grows rapidly with 
time, following the energy input rate and mean hot gas metalicity time 
evolution. It reaches a maximum value after 10 - 20 Myrs, and 
then remains almost constant up to the end of the star forming activity. The 
maximum X-ray luminosity is larger in models with a larger smooth ISM 
component, and matches the value observed in VII Zw403 when the smooth 
component of the ISM amounts to 70\% of the observed HI mass.
Note that we stopped our calculations when the shock fronts reached the 
assumed galactic ISM cut-off position (1.9 kpc). This causes the breaks
in the X-ray curves around 30 Myr and 50 Myr, for the f$_c = 0.9$ and
f$_c = 0.6$ models respectively.  

Note that for these moderate ISM densities and the relevant  
mechanical luminosities (see Figure 4b), the shell of swept-up matter 
becomes radiative after a short time
$\tau_{cool} \approx 2.3 \times 10^4 n_{ISM}^{-0.71} L_{38}^{0.29}$ yr
(Mac Low \& McCray, 1988). However, this is not detected in the available
HI maps. One can then claim that the large-scale shell is photo-ionized
by radiation escaping the central HII region, and thus it should be
most easily 
observed in H$_{\alpha}$ light. However, note that if $\sim 80\%$ of 
the $UV$ photon flux is being used to ionize gas around the central star 
clusters, as it is assumed to be for the totality of clusters 1 and 2 
in VII Zw403  (Lynds et al., 1998), then the rest of the photons 
(N$_{esc} = 0.25 L_{H\alpha} / 1.36 \times 10^{-12}$ 
s$^{-1} \approx 3.25 \times 10^{50}$ s$^{-1}$) would be free to ionize the 
outer ISM. If the number of $UV$ photons trapped within the neutral smooth 
component is negligible, they will ionize the neutral outer shell. Such a 
shell would appear as a 1 kpc diffuse H$_{\alpha}$ emitting feature with 
brightness 
\begin{equation}
      \label{eq.16}
B = \frac{1.36 \times 10^{-12} N_{esc}}{64 \pi R_{sh}^2} \theta^2 \approx
    10^{-17} \, erg \, s^{-1} \, cm^{-2}, 
\end{equation}
where $\theta = 1.5$ arcsec is the averaged seeing. However, this is  
below our detection limit ($8.15 \times 10^{-17} erg \, s^{-1} \, cm^{-2}$), 
and thus the H$_{\alpha}$ emission excess expected at the bubble
outer shell is unlikely to be detected. Approximately an order of
magnitude better sensitivity has to be reached to confirm 
this possible model.

So far it appears that the remnants caused by the two completely different 
recent star-forming histories considered here are able to explain the
observables in VII Zw403. In all of our models however, we have also
calculated the time dependent mixing of the metals freshly ejected by
supernovae with the matter thermally evaporated from the outer shell.
Figure 6 shows the time evolution of the mean inner gas metalicity 
(using oxygen as a tracer, see Silich et al. 2001) predicted for the 
superbubbles in the two extreme
cases able to match the observations of VII Zw403. Both cases show a
superbubble interior metalicity largely different to that of the ISM
detected in the optical regime (Z$_{ISM} = 0.06Z_{\odot}$, see Table 1).
Enormous differences are found in the coeval case which rapidly acquires 
several times Z$_{\odot}$. On the other hand, the continuous star formation, 
although presenting high metalicity values, never surpasses  Z$_{\odot}$.
This result could be used as a discriminator able to discard  
various possible recent histories of star formation in VII Zw403. However,
as it is likely that the swept-up matter in the shell preserves the same
metalicity as the unperturbed ISM (see Tenorio-Tagle, 1996), it will be
difficult to distinguish between the metalicities of two overlaid X-ray 
components in the instantaneous burst case.

A close comparison of the HST observations and our data however has
led us to realize that, although $UV$ photons may escape their 
production centers and HII regions causing the extended diffuse
ionized gas, the mechanical energy from the various star-forming
centers has not had sufficient time to drive a
large-scale remnant. This has not happened, despite the 5 - 10 Myr of 
evolution of the various subgroups in the galactic nucleus. It seems that the
five star-forming centers are sufficiently distant from each other,
that their mechanical energy is still currently being used to build up
individual local shells. The largest H$_{\alpha}$ shell around
star cluster 4 with a radius $\sim$ 250 pc is not yet enclosing,
and not even in contact with the HII regions and smaller shells
produced by the other star forming centers present in the galaxy.

This structure thus invalidates the assumption often made, that a
given mechanical luminosity derived from the H$_{\alpha}$ luminosity 
of a galaxy can be directly used in theoretical models as acting upon 
a smooth medium from the start of the calculation. 

VII Zw403 is clearly indicating that there may be an important leakage
of $UV$ photons out from the centers of star formation, causing an
extended diffuse ionized region around a starburst. However, for the 
mechanical energy of the various well separated centers of star formation 
to start working together in the production of a large-scale remnant, 
a longer time scale is required. A sufficient period needs to allow for
the multiple shock waves to merge and overlap with each
other, clearly implying a longer time the larger the separation 
between subgroups and the denser the ISM may be around the various
star-forming centers.

This time scale is particularly relevant when one tries to match the remnants
produced by the stellar activity in a badly (spatially) resolved or 
distant galaxy, for which one can nevertheless derive a good estimate 
of its H$_{\alpha}$ luminosity.
If for example one will ignore the detailed structure of the ionized
gas in the VII Zw403 and use the mechanical energy input
rate derived directly from the total H$_{\alpha}$ luminosity,
then little difference will be found between the coeval and the continuous 
star formation models. To fit simultaneously the larger H$_{\alpha}$ 
luminosity ($4.7 \times 10^{40}$ erg s$^{-1}$) and the X-ray luminosity 
derived from the PSPC data, both models would require very similar star
cluster masses $(1 - 3) \times 10^6$ M$_{\odot}$, a similar ISM
structure with a large cloud mass filling factor ($\sim 95\%$), and 
a similar superbubble dynamical age (5 - 8 Myr). In both cases the
expanding shell would remain adiabatic when the remnant reaches the 
1 kpc radius. 
That is, if one takes the total H$_{\alpha}$ luminosity to derive
the mechanical energy input rate and the bubble time evolution without 
knowing the ionized gas spatial distribution, one would predict
a very powerful and young remnant. The delay caused by depositing
the derived total mechanical luminosity locally into the different knots, 
clearly changes the outcome. Nevertheless, the high level of energy
deposition (L$_{SB} \approx 2 \times 10^{40}$ erg s$^{-1}$) predicts 
(see Silich \& Tenorio-Tagle, 2001) that the ISM 
of VII Zw403 will be blown away after a few tens of Myrs. 

This leads to another possible interpretation of the observed
X-ray emission, that is coming from the local bubbles generated by the
various stellar subgroups. Take for example the OB associations 1 
and 4 for which we know the total stellar mass, the ionizing flux, 
the age, and the local bubble radius. From these we have derived both 
analytic (Chu \& Mac Low, 1990), and numerical (Silich et al., 2001) 
estimates of the present bubble X-ray luminosity. Note that local 
bubble kinematic ages estimated from the H$_{\alpha}$ radii and 
expansion velocities ($\sim 1$ Myr) are much smaller than the stellar 
cluster ages derived from the massive star isochrones, which are 4 - 6
Myrs (Lynds et al. 1998). This discrepancy is often observed in the
LMC bubbles (Oey \& Smedley, 1998), and most probably is related to
the recent bubble blowout from the host molecular cloud (Silich \& 
Franco, 1999). We have assumed a mean stellar cluster age (5 Myrs) as 
a better indicator of the bubble time evolution. One can then estimate 
the X-ray emission from the local bubbles if the energy input rate 
and the surrounding gas density n$_c$ are known: 
\begin{equation}
      \label{eq.17}
L_x = 10^{36} Z I(\tau) L_{38}^{33/35} n_c^{17/35} t_7^{19/35},
\end{equation}
where  Z is the hot X-ray emitting gas
metalicity, I$(\tau)$ is a dimensionless integral whose value is
close to unity, the energy input rate L$_{38}$, and the evolutionary
time t$_7$, measured in $10^{38}$ erg s$^{-1}$ and $10^7$ yr
units respectively.  We derive the appropriate star cluster mass, 
and its mechanical luminosity from our SB model, and use them to
estimate the surrounding gas density from the standard (Weaver et
al. 1977) bubble model:
\begin{equation}
      \label{eq.18}
n_c = \left(\frac{267 pc}{R_{sh}} \right)^5 L_{38} t_7^3, 
\end{equation}
These results were compared (see table 4) with the numerical 
calculations assuming a time-dependent mechanical energy input rate
and time-dependent hot gas metalicity. Note that the inferred X-ray 
luminosities show a good agreement with the observations.
The large differences between the numerical and analytical estimations 
of the surrounding gas density results from the different mechanical
energy input rates (a constant value for analytical models and
increasing with time rate for numerical calculations).

\section{Conclusion}

Here we have discussed the recent history of star formation and a
possible nature of the diffuse X-ray emission in the nearby BCD 
galaxy VII Zw403. Two possible scenarios of star formation have
been considered: an instantaneous burst, and an extended  
episode of star formation. 

To construct the numerical model we have provided new narrow-band 
observations of VII Zw403 centered on the H$_{\alpha}$ line with
a long exposure time corresponding to the threshold H$_{\alpha}$ flux 
$8.15 \times 10^{-17}$ erg cm$^{-2}$ s$^{-1}$. These observations
reveal a variety of ionized filaments and incomplete shells 
superimposed on the diffuse H$_{\alpha}$ emission that most certainly 
result from the photons leaking out of the main star-forming centers.
The largest feature is the 250 pc broken shell associated with stellar
association 4. The total H$_{\alpha}$ luminosity derived from our 
observations, L$_{H_{\alpha}} = 4.7 \times 10^{40}$ erg s$^{-1}$, 
is much larger than reported before. This has a profound impact on 
the predicted properties of the starburst blown superbubble.

The numerical models based on the HST H$_{\alpha}$ data require
either an instantaneous burst of star formation with a total mass of 
$5 \times 10^5$ M$_{\odot}$, or a star formation episode with a constant
SFR = $4 \times 10^{-3}$ M$_{\odot}$ yr$^{-1}$ lasting 35 Myr.
The models however require radically different structures of the 
galactic ISM and imply very different properties  of the resulting 
remnant.

The best coeval model assumes most of the ISM to be locked up within
high density clouds, and only $\sim 5\%$ of the observed neutral hydrogen 
mass is in the smooth component. The hydrodynamical calculations
also predict the outer shell to be adiabatic after reaching a 1 kpc radius,
and to contribute 50 - 80\% of the observed diffuse X-ray emission.
The bubble evolutionary time is estimated to be $\tau_{dyn} \approx 7$ Myr
when its expansion speed is $\approx 200$ km s$^{-1}$.

The best continuous star formation model requires of much higher density 
in the smooth ISM component, with only $\sim 30\%$ of the HI mass concentrated 
in dense clouds. This leads to a much smaller bubble expansion velocity
(V$_{exp} \approx 35$ km s$^{-1}$), larger evolutionary time 
($\tau_{dyn} \sim 35$ Myr), and a rapid cooling within the outer shell. 
That is, this model predicts a low brightness HI shell surrounding the
diffuse 1 kpc X-ray region. This shell may also show up in ${H\alpha}$ if
exposed to the $UV$ flux from the central cluster. The inner gas
metalicities are also predicted to be very different in these two cases. 

The numerical calculations based on the high energy input rate derived
from our observations require an instantaneous burst or a short
episode of star formation with SFR $\sim 0.1$ M$_{\odot}$ yr$^{-1}$
lasting less than 10 Myr with similar total stellar cluster 
masses $(1 - 3) \times 10^6$ M$_{\odot}$ and most of the ISM ($\sim 95\%$) 
locked up within high density clouds. 
The comparison of the energy input rate derived from our ${H\alpha}$
data with the theoretical limits, implies that the entire ISM and  
metals produced by the current episode of star formation are 
going to be ejected from the galaxy after the coherent 
superbubble is formed. 

It appears that the five main star-forming knots are sufficiently
distant to form a coherent shell in a short time scale, while
keeping their energies blocked within local, spatially separated bubbles.
This provides a time delay that must be considered when developing a
numerical model for the coherent superbubble driven by a number of
young stellar clusters. Numerical calculations show that the
X-ray luminosities from young local bubbles are in a good agreement
with the ROSAT PSPC data. This agreement indicates that the observed 
diffuse component of the X-ray emission may be related to the small 
centrally concentrated bubbles, rather than to the coherent 1 kpc 
structure.  Further observations with the XMM-NEWTON 
observatory is expected to be able to recover the real nature
of the diffuse X-ray emission and the recent history of star
formation in this galaxy.

We thank D. Bomans for his comments and suggestions regarding the X-ray data.
We also thank the anonymous referee for his detailed report that greatly
improved our paper. Finally we also thank Edward Chapin for his
careful reading of the manuscript.

This work has been supported by the Spanish grants PB97-1107 and 
AYA2001-3939, and the Mexico (CONACYT) project 36132-E.

\newpage

\onecolumn

\begin{table}
    \label{tab1} 
      \caption{VII Zw403 observational properties}
       \small
        \begin{flushleft}
\begin{tabular}{|l|l|c|l|} \hline
Parameter &   & Reference & Comments \\
\hline
Distance              & 4.5 Mpc                     &  1   & \\
The galaxy total mass & $2 \times 10^8$ M$_{\odot}$ &  2   & \\
The galaxy ISM mass & ($4 - 7) \times 10^7$ M$_{\odot}$     &  1, 2
                                              & from HI observations \\
The radius of the HI halo & $\sim 1.9$ kpc    & 
                                              & not well known, see text \\ 
The ISM gas velocity     &                    &   &                      \\
dispersion               & $\sim$ 30 km s$^{-1}$   
                                              &  2   & \\
The ISM gas metalicity                       & (0.05 - 0.06)Z$_{\odot}$  
                                              &  3   & \\
The total H$_{\alpha}$ flux & 
          $(195 \pm 4) \times 10^{-13}$ ergs cm$^{-2}$ s$^{-1}$ & our data  
                            &  integrated out to           \\
                            &  &   & the limiting isophote \\
The total H$_{\alpha}$ luminosity    
                       & $1.8 \times 10^{39}$ ergs s$^{-1}$ & 1  &    \\
                       & $4.7 \times 10^{40}$ ergs s$^{-1}$ & our data&  \\
H$_{\alpha}$ shell radii       &  &   &                       \\
\quad - Large                  & $\sim 250$ pc &  our data & around 
                               association 4  \\
\quad - Small                  & $\sim 80$ pc &  1  &  around  
                               association 1        \\
Diffuse X-ray emission:       & $(1.9 - 2.3) \times 10^{38}$ ergs s$^{-1}$ 
                                               &  4, 5 & ROSAT PSPC data \\
\quad - Unresolved core        &  $\sim 85\%$  &  & not confirmed by       \\
\quad - Extended diffuse emission  &  $\sim 15\%$  & & ROSAT HRI data      \\
X-ray flux from the point  &   &   &                                  \\
source                     &$9 \times 10^{-13}$ ergs
                           cm$^{-2}$ s$^{-1}$   & 6   & ROSAT HRI data    \\
\hline
\end{tabular}
\end{flushleft}
\scriptsize{References:                                                     
       1)       Lynds et al., 1998       
       2)       Thuan \& Martin, 1981      
       3)       Schulte-Ladbeck et al., 1999             
       4)       Papaderos et al., 1994 
       5)       Fourniol, 1997 
       6)       Lira et al., 2000
}
\end{table}  


\begin{table}
    \label{tab2} 
      \caption{A 1 kpc bubble parameters for instantaneous burst models}
       \small
        \begin{flushleft}
\begin{tabular} {|l|c|c|c|c|c|c|c|c|} \hline
Model & M$_{SB}$ & f$_c$ &  $\tau_{dyn}$ & $N_{UV}$ & L$_{H\alpha}$ 
      & L$_x$    & Shell state \\
      & \Msol    &   \%  & Myr  & photons s$^{-1}$ 
      & erg s$^{-1}$   &  erg s$^{-1}$ & \\
\hline
  1   & 2 & 3 & 4 & 5 & 6 & 7 & 8        \\
\hline
I5.0\_10 &10$^5$ & 90 & 12.5 &7.4$\times 10^{48}$ & 1.0 $\times 10^{37}$ 
                & 1.2 $\times 10^{37}$ & R \\
I5.0\_40 &10$^5$ & 60 & 21.8 &4.3$\times 10^{47}$ &  5.8 $\times 10^{35}$ 
                & 2.2 $\times 10^{37}$ & R \\ 
I5.0\_70 &10$^5$ & 30 &  29.7 &8.7$\times 10^{46}$ & 1.2 $\times 10^{35}$ 
                & 2.8 $\times 10^{37}$ & R \\ 
I5.7\_5  &$5 \times10^5$  & 95 & 6.7 &9.5$\times 10^{50}$ 
         & 1.3 $\times 10^{39}$ & 4.2 $\times 10^{37}$  & A \\
I5.7\_10 &$5 \times10^5$ & 90 &7.9  &4.3$\times 10^{50}$ 
         & 5.8 $\times 10^{38}$ & 1.6 $\times 10^{38}$  & A \\
I5.7\_40 &$5 \times10^5$ & 60 &12.0 &4.5$\times 10^{49}$ 
         & 6.1 $\times 10^{37}$ & 1.8 $\times 10^{38}$  & R \\
I5.7\_70 &$5 \times10^5$ & 30 &15.5  &1.3$\times 10^{49}$ 
         & 1.7 $\times 10^{37}$ & 2.3 $\times 10^{38}$  & R \\
I6.0\_5 &10$^6$  & 95  & 6.0  &3.5$\times 10^{51}$ & 4.8 $\times 10^{39}$ 
                 & 3.5 $\times 10^{37}$                 & A \\
I6.0\_10 &10$^6$ & 90  & 6.8  &2.0$\times 10^{51}$ & 2.7 $\times 10^{39}$ 
                 & 1.5 $\times 10^{38}$                 & A \\ 
I6.0\_40 &10$^6$ & 60  & 9.8  &2.7$\times 10^{50}$ & 3.7 $\times 10^{38}$ 
                 & 4.5 $\times 10^{38}$                 & R \\ 
I6.0\_70 &10$^6$ & 30  & 12.0 &9.1$\times 10^{49}$ & 1.1 $\times 10^{38}$ 
                 & 5.5$\times 10^{38}$                  & R \\
I7.0\_10 &10$^7$ & 90 & 4.7 &1.5$\times 10^{53}$ & 2.0 $\times 10^{41}$  
                 & 3.3$\times 10^{38}$                  & A \\ 
I7.0\_40 &10$^7$ & 60 & 5.8 &4.4$\times 10^{52}$ & 5.9 $\times 10^{40}$  
                 & 1.7$\times 10^{39}$                  & A \\ 
I7.0\_70 &10$^7$ & 30 & 6.3 &3.0$\times 10^{52}$ & 4.1 $\times 10^{40}$  
                 & 8.1$\times 10^{39}$                  & A \\
\hline
\end{tabular}
\end{flushleft}
\end{table}


\begin{table}
    \label{tab3} 
      \caption{A 1 kpc bubble parameters for continuous star formation models}
       \small
        \begin{flushleft}
\begin{tabular} {|l|c|c|c|c|c|c|c|c|} \hline
Model & M$_{SB}$ & f$_c$ & $\tau_{dyn}$ & $N_{UV}$ & L$_{H\alpha}$ 
      & L$_x$    & Shell state \\
      & \Msol & \% & Myr & photons s$^{-1}$  & erg s$^{-1}$ & erg s$^{-1}$& \\
\hline
  1   & 2 & 3 & 4 & 5 & 6 & 7 & 8                          \\
\hline
C5\_10 &2.1$\times 10^4$ &90 &22.3 &1.3$\times 10^{47}$ &1.6$\times 10^{35}$ 
&1.3$\times 10^{36}$     & R \\
C5\_40 &2.1$\times 10^4$ &60 &50.5 &1.5$\times 10^{45}$ &2.1$\times 10^{33}$  
&1.1$\times 10^{36}$     & R \\
C10\_10 &4.1$\times 10^4$ &90 &20.0 &1.5$\times 10^{48}$ &2.0$\times 10^{36}$ 
&3.6$\times 10^{36}$     & R \\ 
C10\_40 &4.1$\times 10^4$ &60 &34.5 &4.0$\times 10^{46}$ 
        &5.4$\times 10^{34}$  &6.6$\times 10^{36}$     & R \\ 
C10\_70 &4.1$\times 10^4$ &30 &59.0 &1.7$\times 10^{45}$ &2.3$\times 10^{33}$
        & no                                           & R \\
C20\_10 &8.2$\times 10^4$ &90 &19.5 &1.3$\times 10^{51}$ &1.8$\times 10^{39}$ 
        &5.8$\times 10^{36}$                           & R \\ 
C20\_40 &8.2$\times 10^4$ &60 &29.4 &1.7$\times 10^{48}$ &2.3$\times 10^{36}$  
        &1.6$\times 10^{37}$                           & R \\ 
C20\_70 &8.2$\times 10^4$ &30 &37.5 &1.6$\times 10^{47}$ 
        &2.2$\times 10^{35}$  &4.3$\times 10^{37}$     & R \\
C40\_10 &1.6$\times 10^5$ &90 &19.5 &1.3$\times 10^{51}$ &1.8$\times 10^{39}$ 
        &6.3$\times 10^{36}$                           & R \\
C40\_40 &1.6$\times 10^5$ &60 &29.0 &1.3$\times 10^{51}$ &1.8$\times 10^{39}$ 
        &1.9$\times 10^{37}$                           & R \\
C40\_70 &1.6$\times 10^5$ &30 &35.1 &1.3$\times 10^{51}$ &1.8$\times 10^{39}$ 
        &3.2$\times 10^{37}$                           & R \\ 
\hline
\end{tabular}
\end{flushleft}
\end{table}


\begin{table}
    \label{tab4}
      \caption{Parameters from individual bubbles} 
        \begin{flushleft}
\begin{tabular} {|l|l|l|} \hline
       & Association 1 & Association 4 \\
\hline
$UV$ flux & $5 \times 10^{50}$ s$^{-1}$ & $7 \times 10^{50}$ s$^{-1}$
\\
Star cluster age  & 5 Myr                      & 5 Myr                \\
Star cluster mass &$4.7 \times 10^4$ M$_{\odot}$& $6.6 \times 10^4$
M$_{\odot}$ \\
Shell radius  & 79 pc                      & 250 pc                   \\
\hline
                   & Analytic model             & Analytic model  \\
\hline
ISM number density & $1.1 \times 10^3$ cm$^{-3}$& 4.9 cm$^{-3}$   \\
Bubble X-ray luminosity   & $3.5 \times 10^{38}$ Z I($\tau$) erg s$^{-1}$
                  & $3.4 \times 10^{37}$ Z I($\tau$) erg s$^{-1}$  \\
\hline
                   & Numerical model             & Numerical model \\
\hline
ISM number density & 40 cm$^{-3}$                & 0.2 cm$^{-3}$   \\
Bubble X-ray luminosity   & $1.8 \times 10^{38}$ erg s$^{-1}$ 
                          & $2.1 \times 10^{37}$ erg s$^{-1}$     \\
Hot gas mean metalicity   & $Z_O = 1.8 Z_{\odot}$ & $Z_O = 0.6 Z_{\odot}$ \\
\hline
\end{tabular}
\end{flushleft}
\end{table}

\newpage

\begin{figure}
\plotone{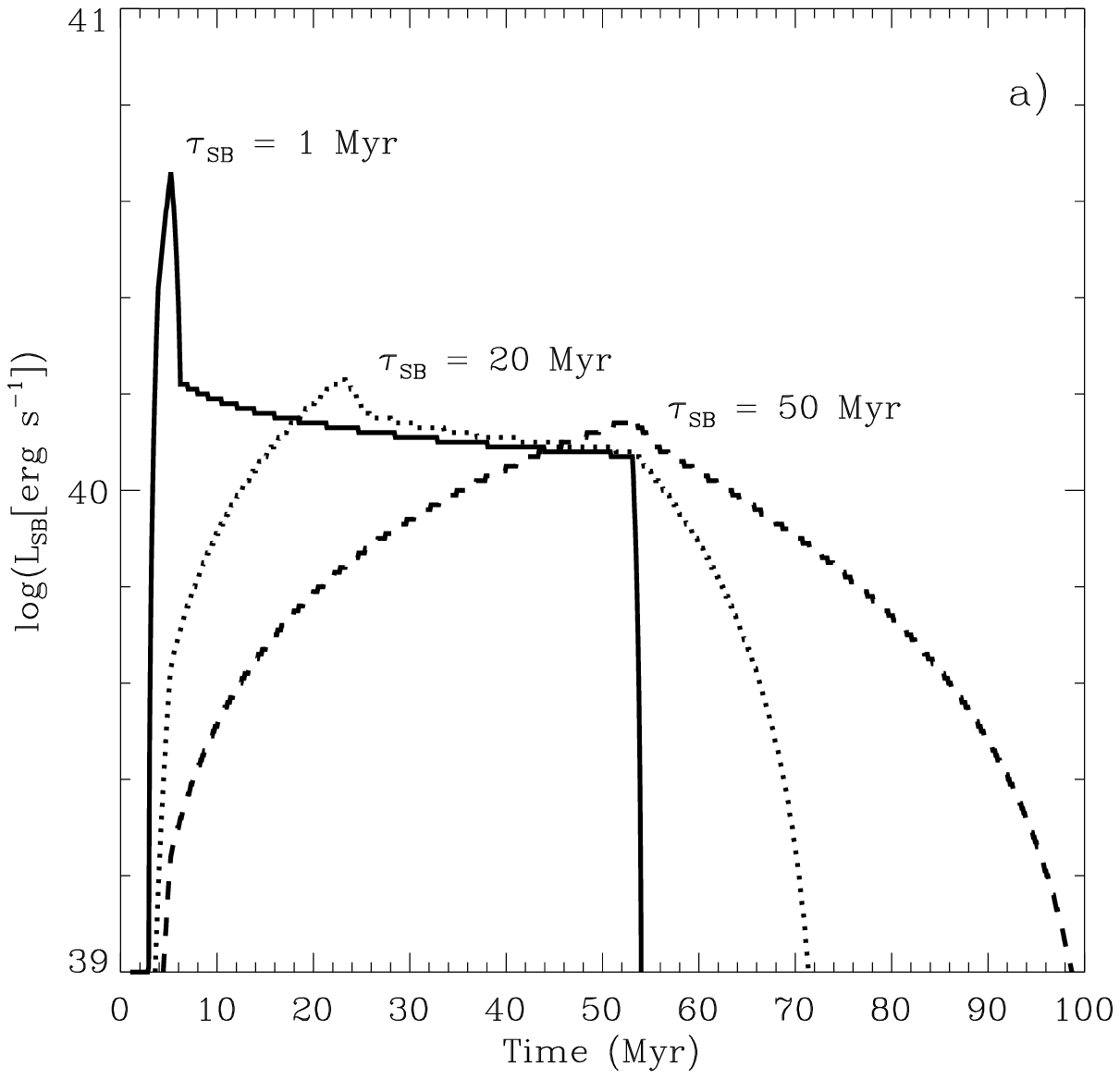} 
\end{figure}

\begin{figure}
\plotone{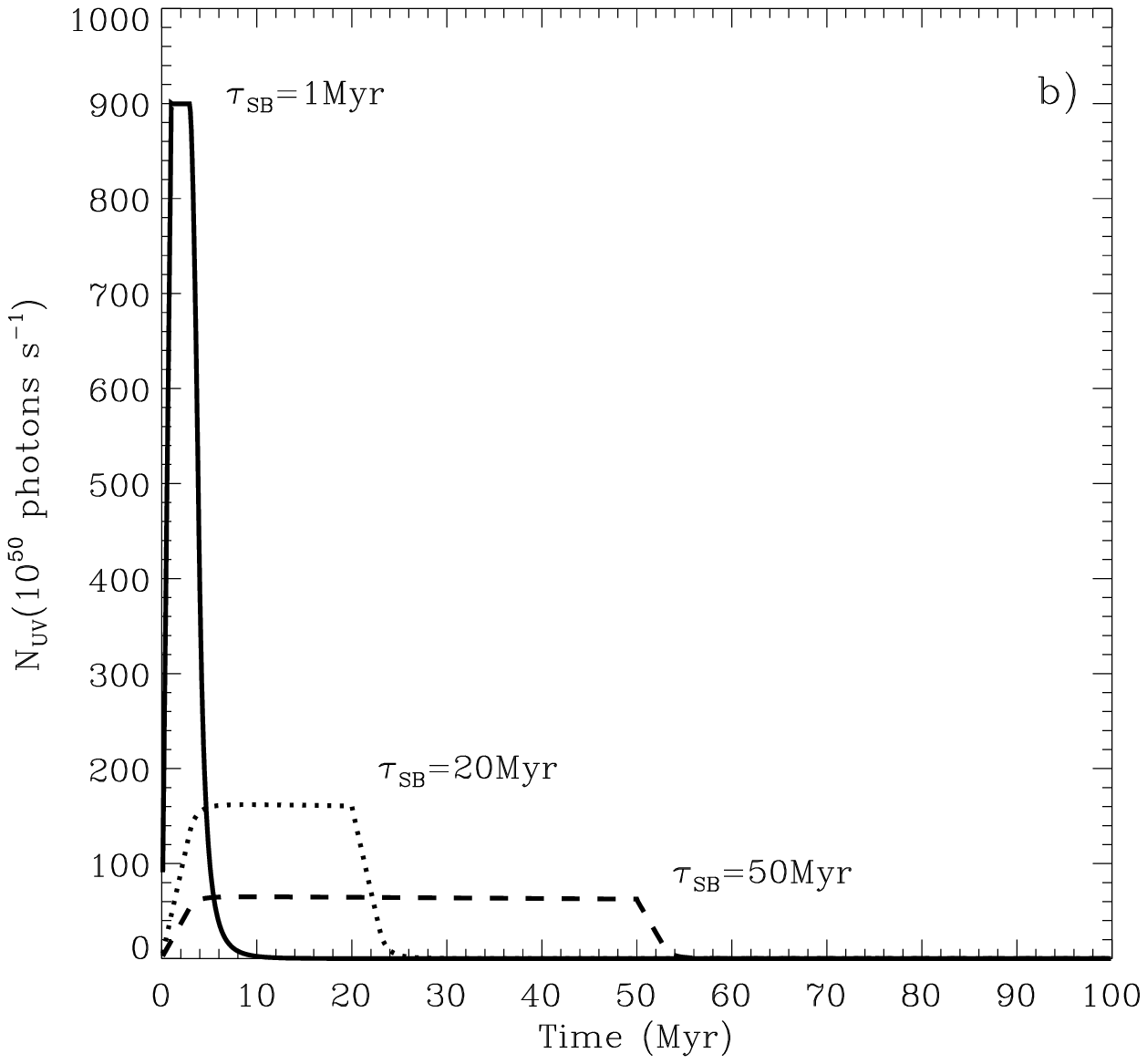} 
\end{figure}

\begin{figure}
\plotone{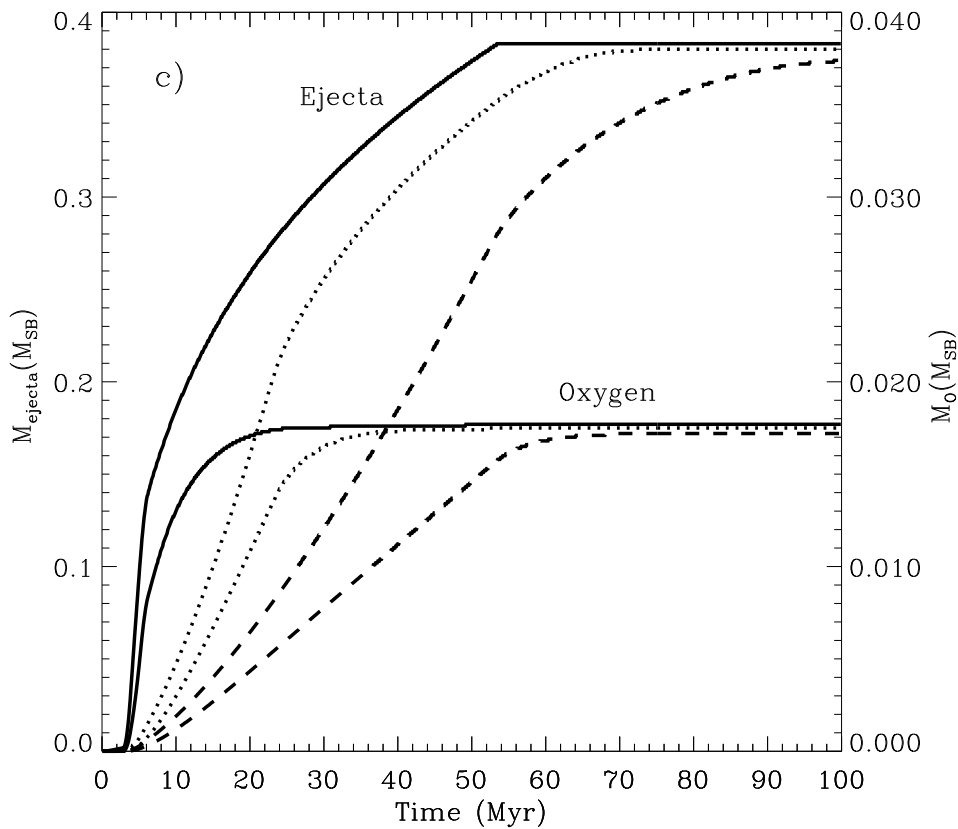}
\caption
{The intrinsic properties of instantaneous and extended starbursts. The panels 
present: the energy input rates (a), the ionizing photon flux (b), and the 
total mass, and the fraction of this in oxygen, ejected by instantaneous and 
extended bursts of star formation (c), all as a function of time. In all the 
panels solid lines represent the instantaneous ($\tau_{SB}$ = 1 Myr)
burst, dotted lines indicate the results for the $\tau_{SB}$ = 20 Myr
and dashed lines 
those for $\tau_{SB}$ = 50 Myr. 
\label{fig1}} 
\end{figure}

\newpage

\begin{figure}
\plotone{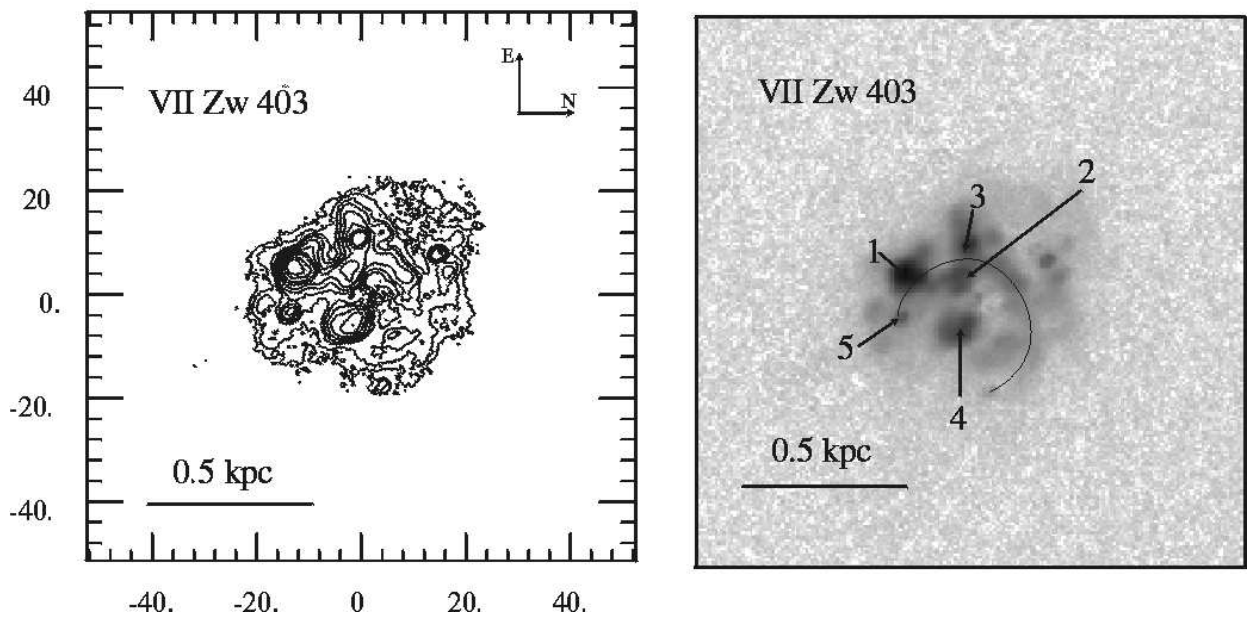}
\caption
{The narrow-band H$_{\alpha}$ image of VII Zw403. The left panel shows
isocontours map of the continuum substracted H$_{\alpha}$ image. The
right panel presents grey scale map of H$_{\alpha}$ emission in logarithmic
scale. The lowest H$_{\alpha}$ isocontour level corresponds to the threshold
value $8.15 \times 10^{-17}$ erg s$^{-1}$ cm$^{-2}$. Isocontours are
equispaced in logaritmic scale.
\label{fig2}}
\end{figure}

\newpage

\begin{figure}
\plotone{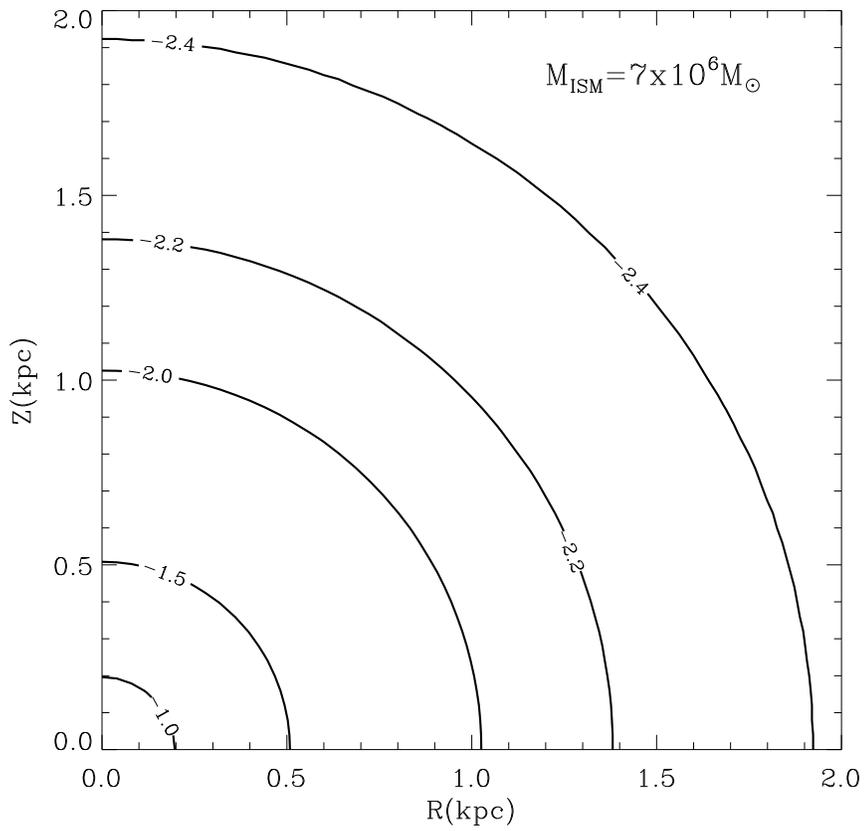}
\caption
{The gas density distribution for models assuming 90\% of the 
ISM mass to be stored in clouds. \label{fig3}}
\end{figure}

\newpage

\begin{figure}
\plotone{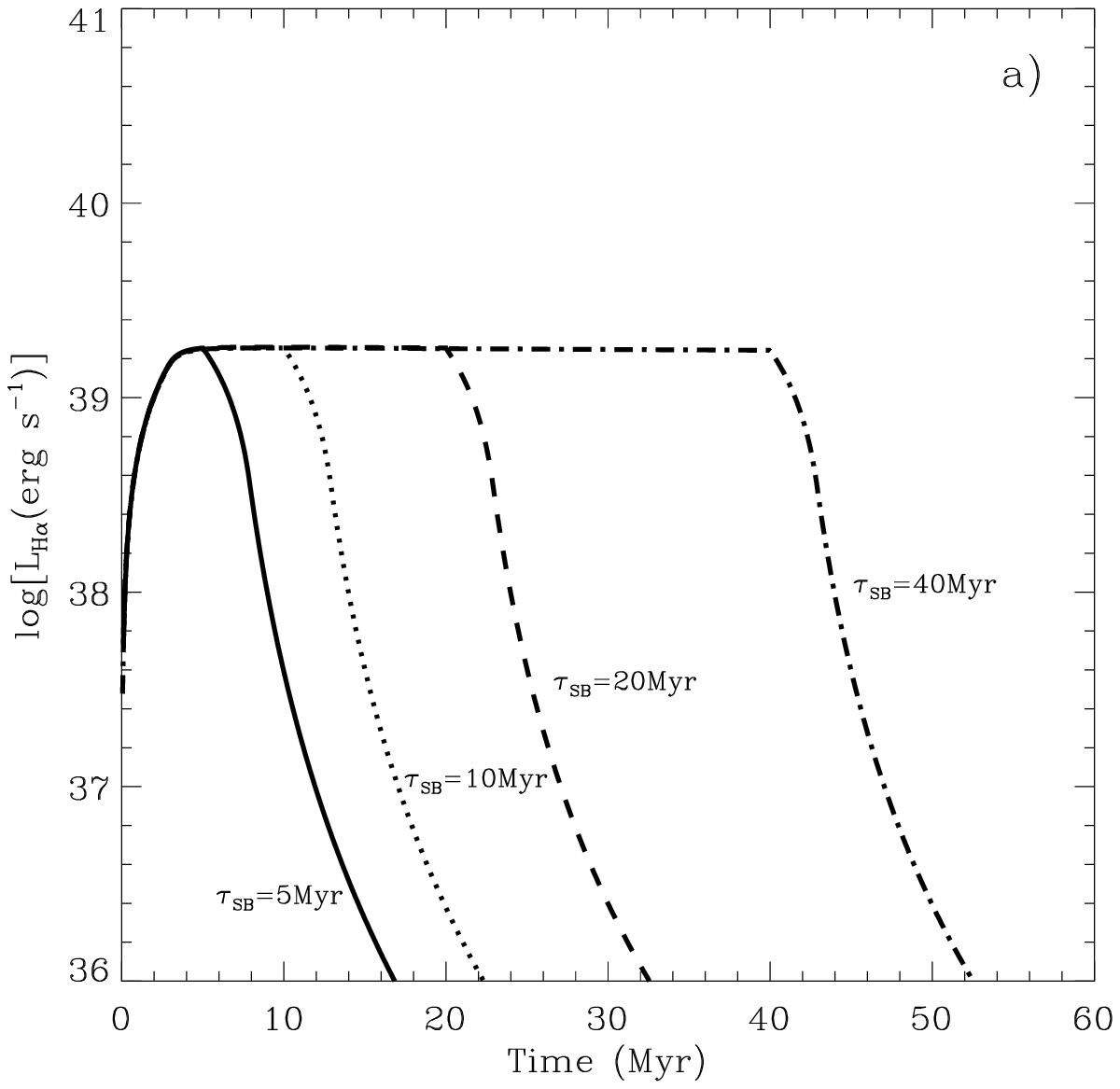}
\end{figure}

\begin{figure}
\plotone{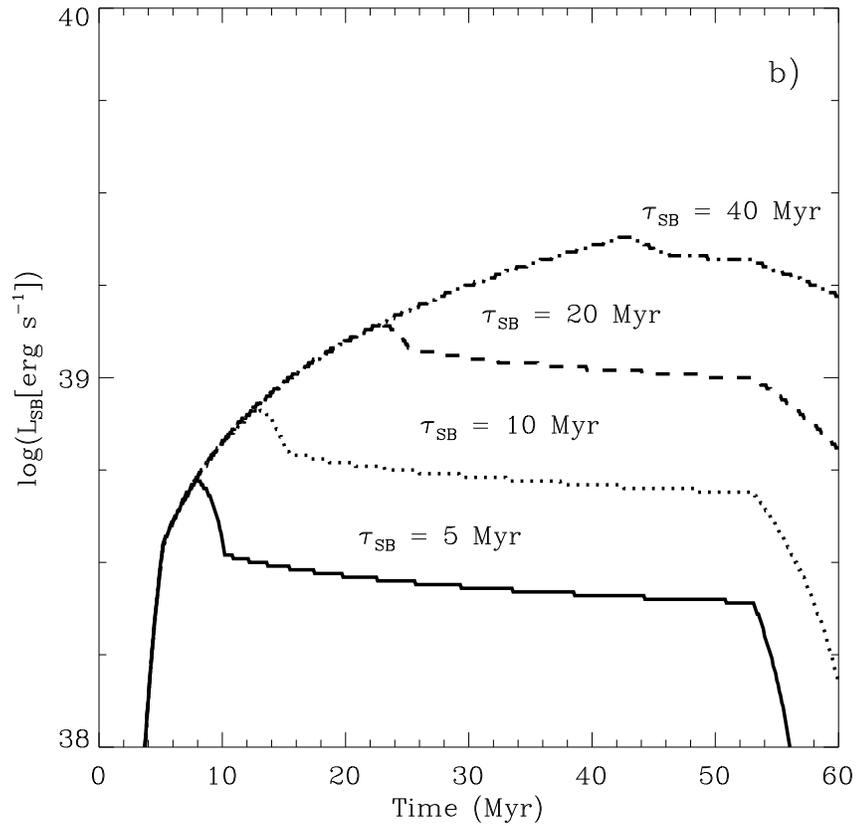}
\caption
{Inferred parameters for VII Zw403. From the observed H$_{\alpha}$ luminosity 
one can infer the SFR and thus the ionizing photon flux. If all these 
quantities remain constant for a $\tau_{SB}$ as shown in panel (a) for the 
central region H$_{\alpha}$ luminosity, one can then infer the run of  the 
mechanical energy input rate (b).\label{fig4}}  
\end{figure}

\newpage

\begin{figure}
\plotone{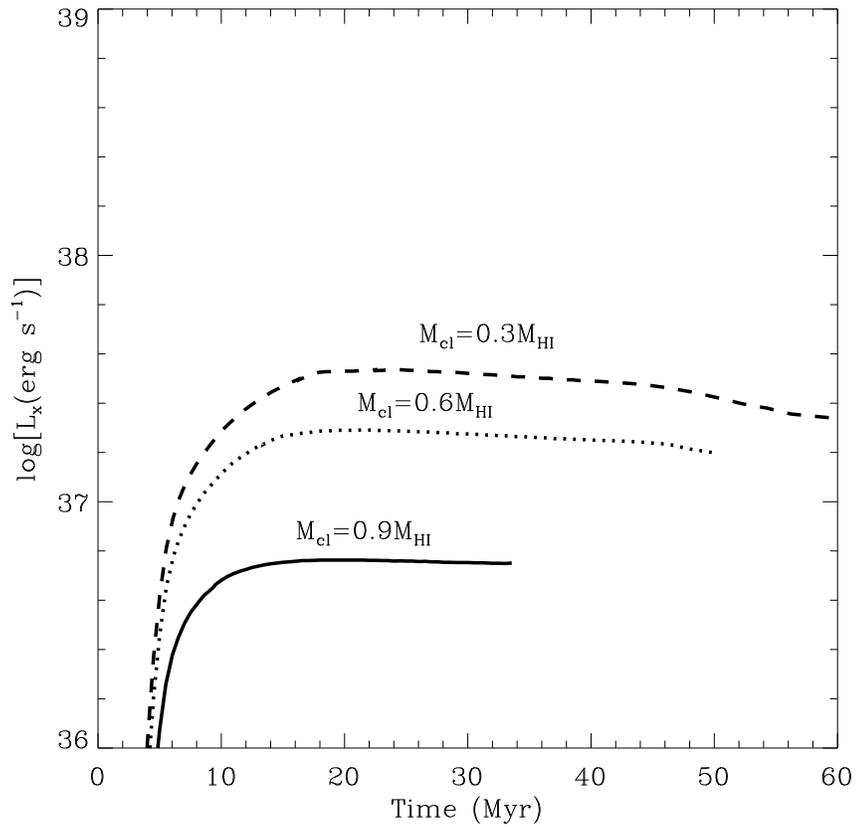}
\caption
{Comparison of the X-ray luminosity (diffuse component) for models with 
$\tau_{SB}$ = 40 Myr and different fractions of the interstellar gas in the 
cloudy component (M$_{cl}$). \label{fig5}}
\end{figure}

\newpage

\begin{figure}
\plotone{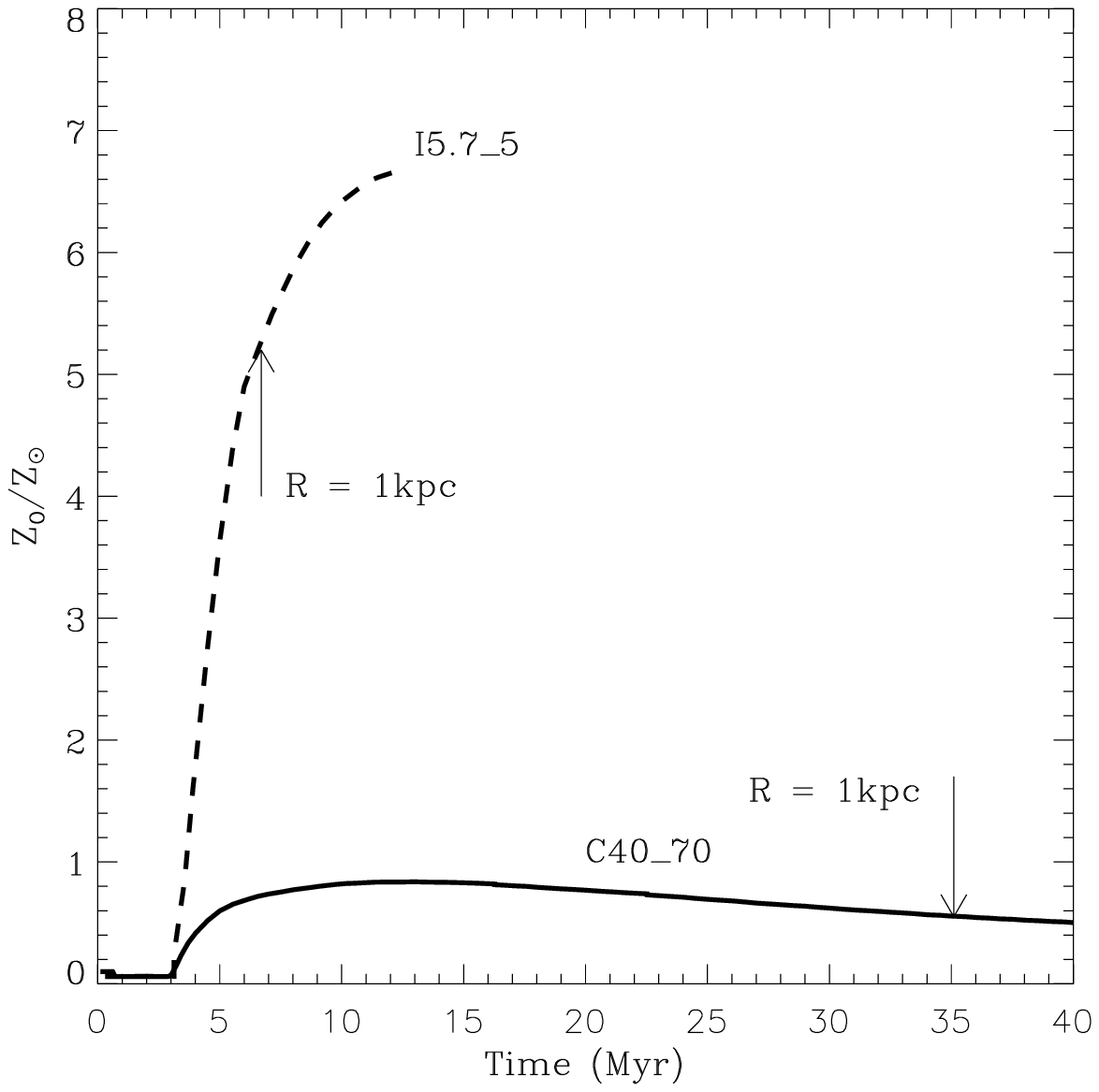}
\caption
{The metalicity of the superbubble as a function of time for two best 
models related with the HST data. Oxygen is used as a tracer, arrows 
indicate metalicities at the time superbubbles reach the 1 kpc
radius. \label{fig6}} 
\end{figure}


\begin{thebibliography}

\bibitem{1} Bisnovatyi-Kogan, G. S. \& Silich, S. A.: 1995, Rev. Mod. Phys. 
            67, 661

\bibitem{2} Beltrametti, M., Tenorio-Tagle, G. \& Yorke, H. W. 1982,
            A\&A, 112, 1

\bibitem{3} Bomans, D. 2001, Rev. Mod. Astroph., 14, 297

\bibitem{4} Burstein, D. \& Heiles, C. 1984, ApJS, 54, 33 

\bibitem{5} Cair\'os, L. M., Caon, N., V\'\i lchez, J. M., 
            Gonz\'alez-P\'erez, N. \& Mu\~noz-Tu\~n\'on, C., 
            2001, ApJ., in press

\bibitem{6} Chiosi, C., Nasi, E. \& Sreenivasan, S. P. 1978, A\&A, 63, 103

\bibitem{7} Chu, Y.-H.\& Mac Low, M.-M., 1990, ApJ, 365, 510

\bibitem{8} Coziol, R., Doyon, R. \& Demers, S. 2001, MNRAS, 325, 1081

\bibitem{9} D'Ercole, A. \& Brighenti, F. 1999,  Mon. Not. Roy. Ast. Soc.,
           309, 941

\bibitem{10} Fourniol, N. 1997, Thesis 'X-ray and optical observations of the
HII galaxies'.

\bibitem{11} Grevesse, N., Noels, A. \& Sauval, A. J. 1996, 
             ASP Conf. Ser., eds. Holt, S. S. \& Sonnebom, G. 99, 117

\bibitem{12} Heckman, T. M., Armus, L. \& Miley, G. K. 1990, ApJS, 74, 833

\bibitem{13} Herbst, W. \& Miller, D. P. 1982, AJ, 87, 1478

\bibitem{14} Leitherer, C. \& Heckman, T. M. 1995, ApJS, 96, 9

\bibitem{15} Lira P., Lawrence A. \& Johnson, R. A. 2000, MNRAS, 319, 17

\bibitem{16} Lynds, R., Tolstoy, E., O'Neil, E. J. Jr. \& Hunter D. A. 1998,
             AJ, 116, 146

\bibitem{17} Maeder, A. 1992, A\&A, 264, 105

\bibitem{18} Mac Low, M.-M. \& McCray, R. 1988, ApJ, 324, 776


\bibitem{20} Oey, M. S. \& Smedley, S. A. 1998, AJ, 116, 1263

\bibitem{21} Oke, J. B. 1990, AJ, 99, 1621

\bibitem{22} Papaderos, P., Fricke, K. J., Thuan, T. X. \& Loose, H.-H. A\&A, 
             1994, 116, L13

\bibitem{23} Schulte-Ladbeck, R. E., Hopp, U., Greggio, L. \& Crone M. M.
             1999, AJ, 118, 2705

\bibitem{24} Shull, J. M. \& Saken, J. M. 1995, ApJ, 444, 663

\bibitem{25} Silich, S. \& Tenorio-Tagle, G. 1998, MNRAS, 299, 249

\bibitem{26} Silich, S. \& Franco, J. 1999, ApJ, 522, 863

\bibitem{27} Silich, S., Tenorio-Tagle, G. Terlevich, R., Terlevich, E. \&
             Netzer, H. 2001, MNRAS, 324, 191

\bibitem{28} Silich, S. \& Tenorio-Tagle, G. 2001, ApJ, 552, 91

\bibitem{29} Stahler, S. W. 1985, ApJ, 293, 207

\bibitem{30} Stothers, R. 1972, ApJ, 175, 431

\bibitem{31} Strickland D. K. \& Stevens I. R. 2000, MNRAS, 314, 511

\bibitem{32} Suchkov, A., Balsara, D., Heckman, T. \& Leitherer, C. 1994,
             ApJ. 430, 511

\bibitem{33} Tenorio-Tagle, G. 1996, AJ, 111, 1641

\bibitem{34} Terlevich, R. 1996, in 11th IAP Astrophysical Meeting, 
             The Interplay Between Massive Star Formation, the ISM and 
             Galaxy Evolution, ed. D. Kunth, B. Guiderdoni, 
             M. Heydari-Malayeri, \& T.X. Thuan (Institut d'Astrophysique,
             Paris), 3

\bibitem{35} Thuan, X. T. \& Martin, G. E. 1981, ApJ, 247, 823

\bibitem{36} Tomisaka, K. \& Bregman, J. N. 1993, PASJ, 45, 513

\bibitem{37} Tully, R. B., Boesgaard, A. M., Dyck, H. M. \& Sckempp, W. V.
             1987, ApJ, 246, 38

\bibitem{38} Weaver R., McCray R., Castor J., Shapiro P. \& Moore R. 1977,
             ApJ, 218, 377

\bibitem{39} Woosley, S. E., Langer, N. \& Weaver, T. A. 1993, ApJ, 411, 823


\end{thebibliography}
\end{document}